\documentclass{sig-alternate}
\usepackage{color}
\usepackage{amsmath}
\usepackage{xcolor}
\usepackage{graphicx}
\usepackage{epsfig}
\usepackage{cite}
\definecolor{gray}{rgb}{0.3,0.3,0.3}
\definecolor{darkblue}{rgb}{0.0,0.5,0.0}

\def\func#1{\textrm{\bf{\sc{#1}}}}

\newcommand{\Oh}[1]{{\mathcal O}\left({#1}\right)}

\def\func#1#2{\mbox{\textrm{\bf{\sc{#1}}}}\ensuremath{(~{#2}~)}}
\def\array#1#2{\mbox{\textrm{\bf{\sc{#1}}}}\ensuremath{[{#2}]}}
\def\arraysize#1{\ensuremath{\size{\mbox{\textrm{\bf{\sc{#1}}}}}}}
\def\var#1{\mbox{\textrm{\bf{\sc{#1}}}}}
\definecolor{gray}{rgb}{0.3,0.3,0.3}

\newcommand{\size}[1]{\ensuremath{\left|#1\right|}}

\newcommand{\Om}[1]{\ensuremath{{\Omega}\left({#1}\right)}}

\newcommand{\Th}[1]{\ensuremath{{\Theta}\left({#1}\right)}}

\makeatletter
   \newcommand\figcaption{\def\@captype{figure}\caption}
   \newcommand\tabcaption{\def\@captype{table}\caption}
\makeatother

\newcommand{\para}[1]{\noindent{\bf{#1}}}

\newcommand{\xif}{{\bf{\em{if~}}}}
\newcommand{\xthen}{{\bf{\em{then~}}}}
\newcommand{\xelse}{{\bf{\em{else~}}}}

\newcommand{\xfor}{{\bf{\em{for~}}}}
\newcommand{\xto}{{\bf{\em{to~}}}}

\newcommand{\xdo}{{\bf{\em{do~}}}}

\newcommand{\xwhile}{{\bf{\em{while~}}}}

\newcommand{\xparallel}{{\bf{\em{parallel~}}}}

\newcommand{\T}{\hspace{0.2cm}}

\newcommand{\xarray}{\mbox{\bf{\em{array}}}}
\newcommand{\xqueue}{\mbox{\bf{\em{queue}}}}

\begin{document}
%
\title{Brief Announcement: An Optimal Level-synchronous Shared-memory Parallel BFS Algorithm with Optimal parallel Prefix-sum Algorithm and its Implications for Energy Consumption\vspace{-0.8cm}}

\vspace{25pt}
{\scriptsize{
\author{
\alignauthor{Jesmin Jahan Tithi, Yonatan Fogel, Rezaul Chowdhury}\\
\affaddr{{Department of Computer Science, Stony Brook University, New York, 11790-4400, USA}}\\
\email{{E-mail:\{rezaul,~yfogel,~jtithi\}@cs.stonybrook.edu}}
}
}
}
\maketitle

\para{Abstract.} We present a work-efficient parallel level-synchronous Breadth First Search (BFS) algorithm for shared-memory architectures which achieves the theoretical lower bound on parallel running time. The optimality holds regardless of the shape of the graph. We also demonstrate the implication of this optimality for the energy consumption of the program empirically. The key idea is to never use more processing cores than necessary to complete the work in any computation step efficiently. We keep rest of the cores idle to save energy and to reduce other resource contentions (e.g., bandwidth, shared caches, etc). Our BFS does not use locks and atomic-instructions and is easily extendible to shared-memory coprocessors. 

\para{Introduction.} Breadth First Search (BFS) is one of the most widely studied graph traversal algorithms with applications spanning from social networks to computational biology.
 Given a graph $G =(V, E)$, with vertex set $V$ ($|V| = n$), edge set $E$ ($|E| = m$) and diameter $D$, a level-synchronous BFS traverses the graph from a given source vertex, $s$ level by level, exploring the first level neighbors of $s$ first, then the second level neighbors and so on until all vertices reachable from the $s$ are explored. A straightforward way to implement a parallel level-synchronous BFS is to explore all vertices at a given level as well as all neighbors of a given vertex in parallel which takes $\Oh{D\log(m+n)}$ time using $O(n+m)$ cores since launching and synchronizing $m+n$ threads requires $\Th{\log(m+n)}$ time. In practice,
the number of available processors $P \ll m + n$. In this paper, we first analyze the theoretical lower bound for a level-synchronous BFS on shared-memory architectures assuming that $k$ threads
can be launched and synchronized in $\Oh{\log k}$ time, and then show that our BFS algorithm is able to archive that optimal lower bound.

\para{Motivation.} It has been observed that for most practical graphs, the number of vertices explored at the beginning and at the end of a BFS exploration is significantly smaller than in the middle levels \cite{Erdos01}. Therefore, using the same amount of computational resources (i.e., number of cores/threads, even caches) at all levels of the BFS is neither economical nor efficient. Indeed, with the increase of the number of active threads synchronization overheads, false sharing, conflict misses and DRAM and CPU energy/power also increase, and without enough work these overheads may dominate the running time. Motivated by this, we have developed a BFS algorithm in which we fix the number of cores based on the amount of work needs to be done at each computation step instead of using all fixed number of cores across the entire BFS. We demonstrate the impact of this choice on theoretical optimality as well as on energy performance on modern multicores.

\para{Algorithm.} Our BFS algorithm (see Figure \ref{parallel-bfs}) uses parallel prefix-\textbf{S}um, \textbf{S}plitting and \textbf{S}canning (hence named {\textbf{\textsc{S3BFS}}) to achieve work-optimality for arbitrary shaped graphs}.
We use the $\var{CurLevelVertices}$ array to store vertices that are going to be explored in a level.
The \textbf{parallel for} loops are implemented using recursive divide-and-conquer where
the recursion stops and switches to an iterative loop when the size becomes $\le \textsc{grainsize}$ and
then each thread executes one of those loops in parallel with others. All major computations in
{\textbf{\textsc{S3BFS}} are performed using \textbf{parallel for}s.

The main loop in lines $9-31$ executes until $\var{CurLevelVertices}$ becomes empty. The degrees of the vertices in $\var{CurLevelVertices}$ are collected in $\var{EdgeSum}$ and then $\var{Parallel-Prefix-Sum}$ \cite{blelloch1990prefix} is used to count the number of edges to be explored in the current level. Then each thread in parallel determines its start location (start vertex and start edge of that vertex) in $\var{CurLevelVertices}$ assuming an (almost) equal division of work. Each active thread explores its assigned segment of edges and stores the newly discovered vertices in a private queue. A thread also marks itself as an owner of a vertex that it stores in its queue for the next level. If multiple threads claims ownership of a vertex, only one of them becomes the owner (benign race). After the exploration, each thread deduplicates its own output queue by keeping only the vertices for which it is the final owner. Next, each thread copies the unique vertices left in its queue to the $\var{CurLevelVertices}$, and the next level of BFS starts.

\para{Analysis.} Let $T_s$ be the running time of the most efficient serial level-synchronous BFS algorithm, let $T_P$ be the running time of such a parallel BFS algorithm on $P \geq 1$ processing cores. 

\textsc{Lower Bound for Parallel BFS:} Since $T_s = \Th{m+n}$, clearly, $T_P = \Om{{{ m+n } \over P}}$. If $W_l$ is the amount of work at level $l$ of a level-synchronous BFS, an optimal algorithm will incur $\Th{\log\min{(P, W_l)}}$ synchronization overhead in that level. Hence, $T_P = \Om{ {{m+n}\over P}+\sum_{l=1}^{D} \log\min{(P, W_l)} }$.\qed

$S3BFS$ achieves the lower bound above as follows. The entire BFS algorithm can be divided into $D$ levels where each level $l$ consists of a constant (say, $c$) number of steps. All of these steps are implemented using \textbf{parallel for} loops. For each such step $i$ with work $W_{l,i}$, we choose $P_{l,i} = \min(P, W_{l, i})$. We distribute the work among $P_{l,i}$ threads in $\Th{\log P_{l,i}}$ parallel time. After that each thread performs $W_{l, i}\over P_{l,i}$ work serially. Hence, the parallel running time of that step is $\Oh{ {W_{l, i}\over P_{l,i}} + \log P_{l,i} }$ which reduces to $\Oh{ {W_{l, i}\over P} + \log P }$ when $P_{l,i} = P$, and to $\Oh{ \log W_{l, i} + 1 }$ when $P_{l,i} = W_{l, i}$. Hence, $\Oh{{W_{l, i}\over P} + \log \min(P, W_{l, i}) + 1}$ is the parallel running time covering both cases.

Therefore, $T_P = \sum_{l=1}^D {\sum_{i=1}^c}$ $\Oh{{W_{l, i}\over P} + \log \min(P, W_{l, i}) + 1}$
$= \Oh{{\sum_{l=1}^D {\sum_{i=1}^c}W_{l, i}\over P} + \sum_{l=1}^D {\sum_{i+1}^c{ \log \min(P, W_{l, i}) } } + cD}$
$= \Oh{ {{m+n}\over P} + \sum_{l=1}^D{ \log \min(P, W_{l}) } }$, where $W_{l} = \sum_{i = 1}^{c}{(W_{l, i})}$.

The $\var{Find-StartExpPoint}$ in Line $17$ is used to find the starting exploration location of each
active thread. Instead of binary search we use the following approach so that it can be executed as
a \textbf{parallel for}. For each entry of the $\var{EdgeSum}$ array, in parallel we first determine a thread id, $t_s$ that can start at that location (i.e., a thread that should start exploring an edge connected to that node) and a thread id, $t_e$ that could possibly end at that node. Note that this range will be empty for all except $\le {P_{l,i}}$ of them. For each nonempty range, we write this node's position as the corresponding thread's start location. 

\para{Experimental Results.}
We implemented our work-sensitive \textsc{S3BFS} using Intel's {\tt{Cilk Plus}}. To show the energy benefits of choosing an optimal number of cores at each computation step, we profiled \textsc{S3BFS} while running on the wikipedia graph with $1000$ random sources. We repeated each run $30$ times with each source, and the entire experiment $4$ times, and took the average. We ran a work-insensitive version of S3BFS in which the maximum number of cores were used for the entire computation. Figure \ref{Scalability}$(a)$ shows a performance comparison of these two versions.
In this figure a ratio of $> 1$ means that the work-sensitive $S3BFS$ is doing better, otherwise, the work-insensitive version is doing better. Since the first two BFS levels of the wikipedia graph typically have very few vertices and edges, the work-sensitive version should win which is, indeed, the case in the Figure. Since in level $3$ the amount of work at each step is more than the maximum number of cores in our multicore machine ($16$-core Xeon Sandybridge), the ratios drop slightly bellow $1$. Even then interestingly the ratio of DRAM energy consumption is still $>1$.
This shows potential energy benefit of \textsc{S3BFS} especially while running on thousands of cores.

Figure \ref{Scalability}$(b)$ shows scalability of \textsc{S3BFS} on a $61$-core Intel Xeon Phi architecture in its native mode on different real-world and synthetic graphs. It shows that S3BFS scales almost linearly till $120$ threads
on Xeon Phi for large enough graphs.

\begin{figure}[t!]
\vspace{-0.2cm}
\begin{minipage}{.47\textwidth}
\vspace{-0.1cm}
\framebox{
\begin{minipage}{\textwidth}
{\scriptsize{
\noindent $\func{Parallel-BFS}{s, ~P}$ $s$ is the source vertex from which distance is calculated. $P$ is the maximum number of processors to use. Returns $\array{d}{0:n-1}$ which represents the distance from $s$ to each vertex.                    
    \noindent
    \vspace{-0.06cm}
    \begin{enumerate}
    \vspace{-0.1cm} \item $P_l \gets \func{Min}{P, ~n}$
    \vspace{-0.1cm}  \item \xparallel \xfor $u \gets 0$ \xto $n-1$ \xdo{\textcolor{blue}{ //\textbf{for} grainsize = $n/P_l$.}}
     \vspace{-0.1cm} \item \T $\array{d}{u} \gets \infty$, $\array{Owner}{u} \gets P$ \textcolor{blue}{ //{initialize distance \& owner.}}
     \vspace{-0.1cm}  \item $\array{d}{s} \gets 0$, $\array{Owner}{s} \gets 0$ \textcolor{blue}{ //initialize source values.}
     \vspace{-0.1cm}  \item $\var{CurLevelVertices} \gets \xarray[0:n-1]$ \textcolor{blue}{ //current level nodes.}
      \vspace{-0.1cm}  \item $\var{EdgeSum} \gets \xarray[0:n-1]$\textcolor{blue}{ //holds prefixsum of degrees.}
     \vspace{-0.1cm}  \item $\array{CurLevelVertices}{0} \gets s$ \textcolor{blue}{ //insert source vertex.}
     \vspace{-0.1cm}  \item $\var{l} \gets 0$, $P_l \gets 1$ \textcolor{blue}{ //initial level and number of threads.}
     \vspace{-0.1cm}   \item \xwhile $\var{$n_l$} \gets \arraysize{CurLevelVertices} > 0$ \xdo \textcolor{blue}{ //any vertex left.}
     \vspace{-0.1cm}   \item \T $\var{l} \gets \var{l} + 1$ \textcolor{blue}{ //$n_l$ = \#vertices at level $l$.}
     \vspace{-0.1cm}   \item \T $\var{$P_l$} \gets min(P, n_l)${\textcolor{blue}{ //$P_l$ chosen based on max work.}}
     \vspace{-0.1cm}   \item \T \xparallel \xfor $u \gets 0$ \xto $n_l-1$ \xdo \textcolor{blue}{ //\textbf{for} grainsize = $n_l/P_l$.}
     \vspace{-0.1cm}   \item \T \T $\array{EdgeSum}{u} \gets Degree[\array{CurLevelVertices}{u}]$
     \vspace{-0.1cm}   \item \T $\func{Parallel-Prefix-Sum}{\var{EdgeSum}, n_l, P_l}$ \textcolor{blue}{ //degree sum.}
     \vspace{-0.1cm}   \item \T $e_l \gets \array{EdgeSum}{n_l-1}$ \textcolor{blue}{//total edges going to be explored.}
      \vspace{-0.1cm}  \item \T $P_l \gets \func{Min}{P, e_l}$\label{reduce-p}{\textcolor{blue}{ //$P_l$ chosen based on max work.}}

     \textcolor{blue}{ //Find starting exploration point for each active thread.}

      \vspace{-0.2cm} \item \T $\var{StartExpPoint} \gets \func{Find-StartExpPoint}{\var{EdgeSum}, ~e_l, ~P_l}$\label{skip-binary-search}

      \textcolor{blue}{ //Main Exploration with optimal number of threads.}

       \vspace{-0.2cm} \item \T $\var{Q}$ $\gets$ {\tiny EXPLORE-VERTEX} ($\var{CurLevelVertices}, \var{EdgeSum}$, $\var{StartExpPoint}, \var{d}$, Neighbor, Degree, $e_l$, $P_l$, $\var{l}$)

       {\textcolor{blue}{ //deduplication (keep single copy of a vertex).}}
        \vspace{-0.1cm}\item \T $\var{Sizes} \gets \xarray[0:P_l-1]$ {\textcolor{blue}{//stores output Q sizes.}}
        \vspace{-0.1cm}\item \T \xparallel \xfor $i \gets 0$ \xto $P_l-1$ \xdo {\textcolor{blue}{ //\textbf{for} grainsize = 1.}}
        \vspace{-0.1cm}\item \T \T $\var{Q-New} \gets \xqueue$ {\textcolor{blue}{ //temporary local queue.}}
        \vspace{-0.1cm}\item \T \T \T \xfor $v$ $\in$ $\array{Q}{i}$ \xdo \xif $\array{Owner}{v} = i$ \xthen $\var{Q-New}.\func{Enqueue}{v}$
        \vspace{-0.3cm}\item \T \T $\array{Q}{i} \gets \var{Q-New}$, $\array{Sizes}{i} \gets \size{\array{Q}{i}}$
        \vspace{-0.1cm}\item \T $\func{Parallel-Prefix-Sum}{\var{Sizes}, P_l, P}$

        {\textcolor{blue}{ //linearization (copy from distributed Q to a linear Q).}}
        \vspace{-0.1cm}\item \T $\var{CurLevelVertices} \gets \xarray[0:\array{Sizes}{P_l-1}]$
        \vspace{-0.1cm}\item \T \xparallel \xfor $i \gets 0$ \xto $P_l-1$ \xdo
        \vspace{-0.1cm}\item \T \T \xif $i = 0$ \xthen $\var{Offset} \gets 0$ \xelse $\var{Offset} \gets \array{Sizes}{i-1}$
        \vspace{-0.1cm}\item \T \T \xfor $j \gets \var{Offset}$ \xto $\var{Offset} + \size{\array{Q}{i}}$ \xdo
        \vspace{-0.1cm}\item \T \T \T $\array{CurLevelVertices}{{j}} \gets \array{Q}{i}.\func{Dequeue}{}$
    \end{enumerate}
    \vspace{-0.1cm}
}
}
\end{minipage}
}
\end{minipage}
\vspace{-.1cm}
\caption{{S3BFS: level-synchronous parallel breadth-first search.}}
\label{parallel-bfs}
\end{figure}
\begin{figure}[h!]

	\centering
     {\includegraphics[width=0.233\textwidth, clip = true]{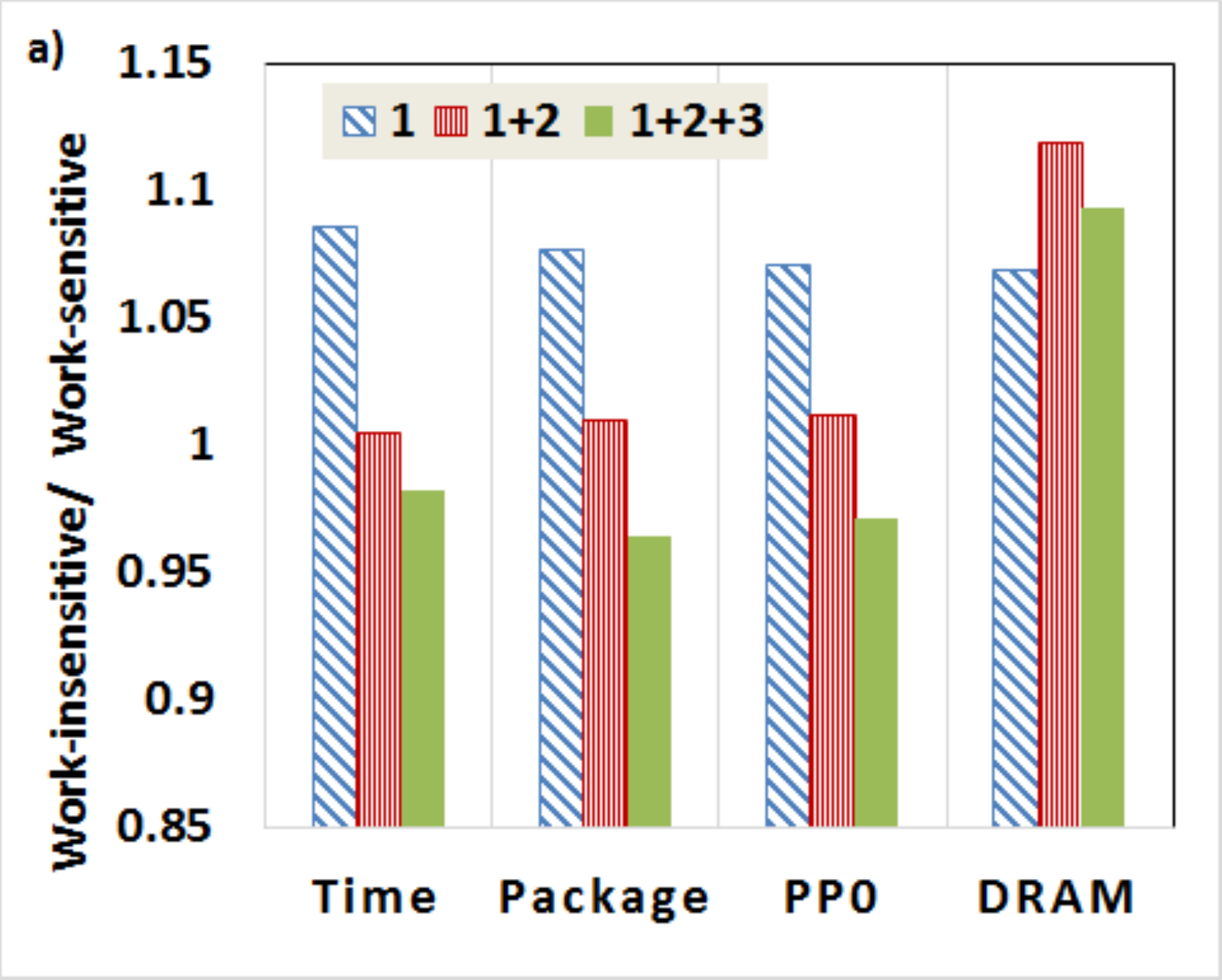}} \hfill
       {\includegraphics[width=0.233\textwidth, clip = true]{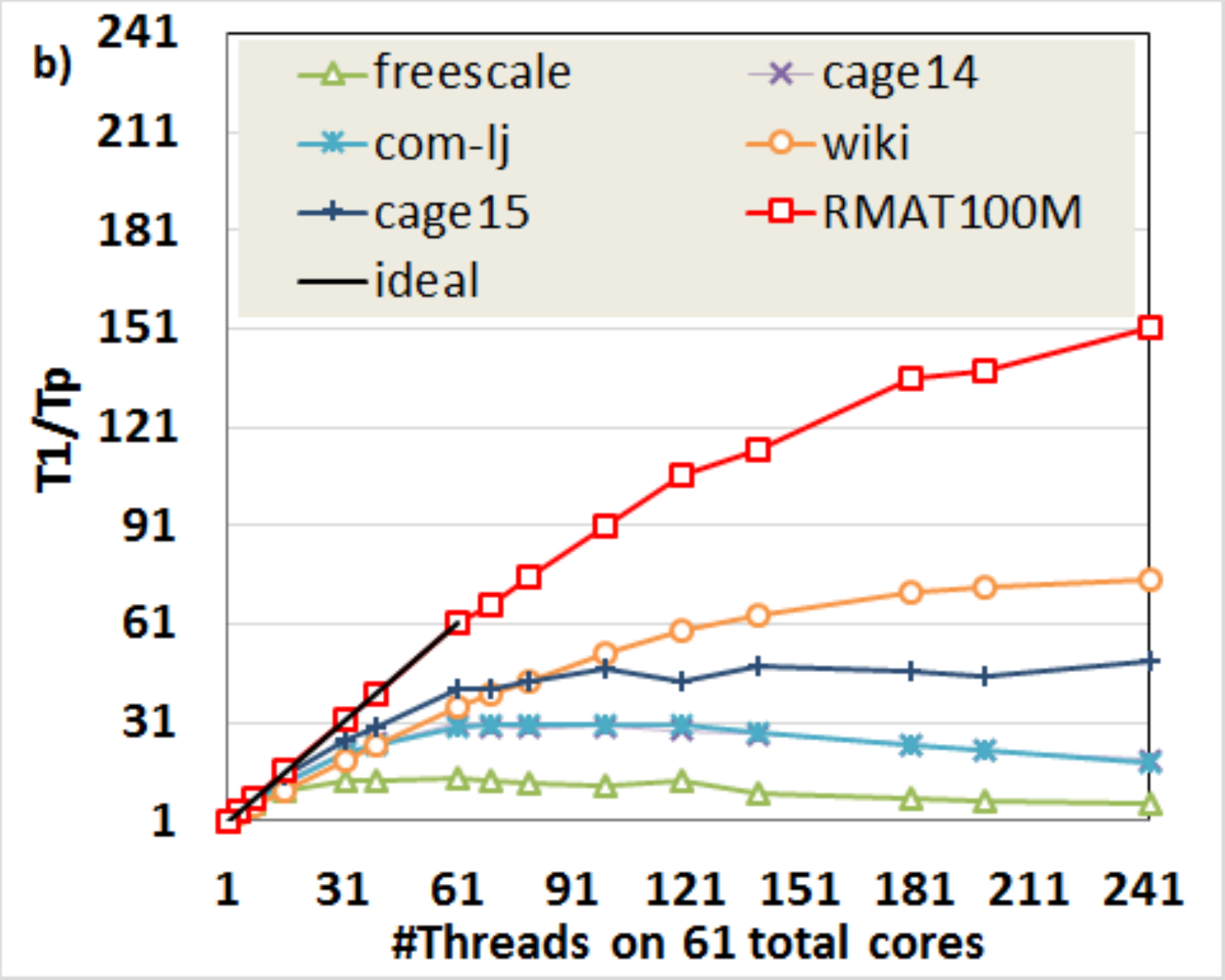}} \hfill
     \vspace{-0.5cm}
  \caption{$(a)$ Energy consumption (Package = Socket (CPUs), PP0 = Power Plane0), $(b)$ Scalability on Xeon Phi.}
\label{Scalability}
\vspace{-0.5cm}
\end{figure}

\para{Conclusion.}
We present a theoretically optimal level-synchronous parallel BFS algorithm which achieves optimality and reduces energy consumption by actively controlling the number of cores used in each computation step. We empirically show that for levels where the amount of work is significantly less than the available cores, this approach reduces energy and power considerably. This algorithm can bring more energy benefits while running BFS on thousands of processing cores and can be optimized with all known optimizations (e.g., direction optimizations). We believe similar techniques can be used for distributed and coprocessor settings, which probably can bring even more energy savings.

\newpage
\bibliographystyle{plain}
\bibliography{bib/briefAnnouncement}

\end{document}